\documentstyle[epsf]{mn}

\newif\ifAMStwofonts

\def\spose#1{\hbox to 0pt{#1\hss}}
\def\lta{\mathrel{\spose{\lower 3pt\hbox{$\mathchar"218$}}
     \raise 2.0pt\hbox{$\mathchar"13C$}}}
\def\gta{\mathrel{\spose{\lower 3pt\hbox{$\mathchar"218$}}
     \raise 2.0pt\hbox{$\mathchar"13E$}}}

\let\gtrsim=\gta

\def\msun{{\rm\,M_\odot}}

\mathchardef\star="313F

\def\gtrsim{\mathrel{\hbox{\rlap{\hbox{\lower4pt\hbox{$\sim$}}}\hbox{$>$}}}}
\def\mmatrix#1{\null\,\vcenter{\normalbaselines
    \ialign{\hfil$##$\hfil\,\,&&\hbox to1pt{\mathstrut\hss\vrule\hss}\,\,\hfil$##$\hfil\crcr
      \mathstrut\crcr\noalign{\kern-\baselineskip}
      #1\crcr\mathstrut\crcr\noalign{\kern-\baselineskip}}}\,}

\newdimen\colsize
\colsize=\textwidth
\divide\colsize by 2
\advance\colsize-12.1pt

\ifoldfss
  \ifCUPmtlplainloaded \else
    \NewTextAlphabet{textbfit} {cmbxti10} {}
    \NewTextAlphabet{textbfss} {cmssbx10} {}
    \NewMathAlphabet{mathbfit} {cmbxti10} {} 
    \NewMathAlphabet{mathbfss} {cmssbx10} {} 
  \fi
  \ifAMStwofonts
    \ifCUPmtlplainloaded \else
      \NewSymbolFont{upmath} {eurm10}
      \NewSymbolFont{AMSa} {msam10}
      \NewMathSymbol{\upi}     {0}{upmath}{19}
      \NewMathSymbol{\umu}     {0}{upmath}{16}
      \NewMathSymbol{\upartial}{0}{upmath}{40}
      \NewMathSymbol{\leqslant}{3}{AMSa}{36}
      \NewMathSymbol{\geqslant}{3}{AMSa}{3E}

       \let\ge=\geqslant
    \fi
  \fi
\fi 

\ifnfssone
  \newmathalphabet{\mathit}
  \addtoversion{normal}{\mathit}{cmr}{m}{it}
  \addtoversion{bold}{\mathit}{cmr}{bx}{it}
  \newmathalphabet{\mathbfit} 
  \addtoversion{normal}{\mathbfit}{cmr}{bx}{it}
  \addtoversion{bold}{\mathbfit}{cmr}{bx}{it}
  \newmathalphabet{\mathbfss} 
  \addtoversion{normal}{\mathbfss}{cmss}{bx}{n}
  \addtoversion{bold}{\mathbfss}{cmss}{bx}{n}
  \ifAMStwofonts
    \ifCUPmtlplainloaded \else
      \UseAMStwoboldmath
      \makeatletter
      \new@mathgroup\upmath@group
      \define@mathgroup\mv@normal\upmath@group{eur}{m}{n}
      \define@mathgroup\mv@bold\upmath@group{eur}{b}{n}
      \edef\UPM{\hexnumber\upmath@group}
      \new@mathgroup\amsa@group
      \define@mathgroup\mv@normal\amsa@group{msa}{m}{n}
      \define@mathgroup\mv@bold\amsa@group{msa}{m}{n}
      \edef\AMSa{\hexnumber\amsa@group}
      \makeatother
      \mathchardef\upi="0\UPM19
      \mathchardef\umu="0\UPM16
      \mathchardef\upartial="0\UPM40
      \mathchardef\leqslant="3\AMSa36
      \mathchardef\geqslant="3\AMSa3E

       \let\ge=\geqslant
    \fi
  \fi
\fi 

\ifnfsstwo
  \DeclareMathAlphabet{\mathbfit}{OT1}{cmr}{bx}{it}
  \SetMathAlphabet\mathbfit{bold}{OT1}{cmr}{bx}{it}
  \DeclareMathAlphabet{\mathbfss}{OT1}{cmss}{bx}{n}
  \SetMathAlphabet\mathbfss{bold}{OT1}{cmss}{bx}{n}
  \ifAMStwofonts
    \ifCUPmtlplainloaded \else
      \DeclareSymbolFont{UPM}{U}{eur}{m}{n}
      \SetSymbolFont{UPM}{bold}{U}{eur}{b}{n}
      \DeclareSymbolFont{AMSa}{U}{msa}{m}{n}
      \DeclareMathSymbol{\upi}{0}{UPM}{"19}
      \DeclareMathSymbol{\umu}{0}{UPM}{"16}
      \DeclareMathSymbol{\upartial}{0}{UPM}{"40}
      \DeclareMathSymbol{\leqslant}{3}{AMSa}{"36}
      \DeclareMathSymbol{\geqslant}{3}{AMSa}{"3E}

       \let\ge=\geqslant
    \fi
  \fi
\fi 

\ifCUPmtlplainloaded \else
  \ifAMStwofonts \else 
    \def\upi{\pi}
    \def\umu{\mu}
    \def\upartial{\partial}
  \fi
\fi

\title{Fluctuations in finite $N$ equilibrium stellar systems}

\author[Martin D. Weinberg]{Martin D. Weinberg\thanks{Alfred P. Sloan
  Foundation Fellow.\newline e-mail: weinberg@phast.umass.edu}\\
  Department of Physics and Astronomy, University of Massachusetts,
  Amherst, MA 01003-4525, USA}

\date{}

\pagerange{\pageref{firstpage}--\pageref{lastpage}}

\pubyear{1997}

\begin{document}

\maketitle

\label{firstpage}

\begin{abstract}
  Gravitational amplification of Poisson noise in stellar systems is
  important on large scales.  For example, it increases the dipole
  noise power by roughly a factor of six and the quadrupole noise by
  50\% for a King model profile.  The dipole noise is amplified by a
  factor of fifteen for the core-free Hernquist model.  The
  predictions are computed using the dressed-particle formalism of
  Rostoker \& Rosenbluth (1960) and are demonstrated by n-body
  simulation.  
  
  This result implies that a collisionless n-body simulation is
  impossible; The fluctuation noise which causes relaxation is an
  intrinic part of self gravity.  In other words, eliminating {\it
    two-body} relaxation does not eliminate relaxation altogether.
  
  Applied to dark matter halos of disk galaxies, particle numbers of
  at least $10^6$ will be necessary to suppress this noise at a level
  that does not dominate or significantly affect the disk response.
  Conversely, halos are most likely far from phase-mixed equilibrium
  and the resulting noise spectrum may seed or excite observed
  structure such as warps, spiral arms and bars.  For example,
  discreteness noise in the halo, similar to that due to a population
  of $10^6M_\odot$ black holes can produce observable warping and
  possibly excite or seed other disk structure.
\end{abstract}


\section{Introduction}

Stellar systems are finite number systems by nature.  Because the
characteristic relaxation time is so long in many cases of
astronomical interest, the continuum limit is a good approximation and
the collisionless Boltzmann equation (CBE) governs the evolution.
Because the CBE is difficult to solve in any generality, many
researchers have turned to n-body simulation as a primary source of
insight.  For a modest number of particles, however, the naive n-body
problem is not a simulation of the CBE but of the {\it collisional}
Boltzmann equation.  Enormous effort has been applied to manipulating
the interparticle force to reduce the intrinsic relaxation to produce
a near-collisionless solution.  The most commonly used technique
smoothes the particle over a finite size region leading to the {\it
  softened} the point mass potential: $G\mu/r\rightarrow
G\mu/\sqrt{r^2+a^2}$.  As long as the smoothing size is not larger
than the mean interparticle spacing, intuitively, there should be no
significant change in the results.

Unfortunately, two-body interactions are only part of the story.
Poisson fluctuations excite structure at all scales in the system.
Many simulations are optimized to resolve large-scale features but the
relaxation is enhanced by large-scale collective excitations on these
same scales (Weinberg 1993\nocite{Wein:93}).  One can think of this
excitation as the projection Poisson noise on the modal spectrum of
the stellar system.  This is the same modal spectrum responsible for
producing a response by a perturber of interest, such as a galaxy
reacting to orbiting or passing companion.  Therefore, a collisionless
solution is physically impossible without irrevocably changing the
dynamics of the system under study.  In other words, one suppresses
the global part of the relaxation at the risk of throwing out the
physics responsible for the evolution one is studying.  There is no
other recourse but large numbers of particles.

For example, this work was motivated by n-body experiments with a thin
disk in a live halo.  The observed fluctuations vertical in force at
the disk plane were larger than predicted for Poisson fluctuations for
n-body simulations of halos with $10^5$ particles using the SCF scheme
(e.g.  Clutton-Brock 1972, 1973, Hernquist \& Ostriker
1992\nocite{Clut:72,Clut:73,HeOs:92}).  However, real galactic halos
are certainly not smooth and contain gas clouds, star clusters, dwarf
galaxies, stellar streams, and possibly as of yet undetected massive
objects such as $10^6M_\odot$ black holes (e.g. Lacey \& Ostriker
1985\nocite{LaOs:85}).  All of these can contribute to correlated
fluctuations at large scales leading to warped disks and other
possibly observable distortions (see Weinberg 1997\nocite{Wein:97}).

In this paper, I describe the expected amplitude of noise-generated
fluctuations in a spherical equilibrium stellar system including
self-gravity.  The main result is the power spectrum of fluctuations
generated by long-range correlations of particles moving their own
gravitational field.  This is computed using the polarization cloud
method developed by Rostoker \& Rosenbluth (1960\nocite{RoRo:60}) for
plasma physics.  The same approach can be used for any regular system
(see Nelson \& Tremaine 1997\nocite{NeTr:97} for a general
discussion). The power at very small scales will be Poisson, but at
large scales, it will be modified by the global gravitational
response.  The basic results developed in \S\ref{sec:derive} show that
this might have been predicted a priori and is applied to a few
standard spherical models in \S\ref{sec:examples} and the predictions
corroborated by n-body simulation. We conclude in \S\ref{sec:conclu}.

\section{Derivation}
\label{sec:derive}

\subsection{Basic approach}
\label{sec:cbe}

Since we are concerned with perturbations to an equilibrium, we may
solve the linearized collisionless Boltzmann equation (LCBE):
\begin{equation}
  \label{eq:lcbe}
  {\partial f_1\over\partial t} + 
  {\partial f_1\over\partial{\bf w}}\cdot {\partial H_o\over\partial{\bf I}} -
  {\partial f_o\over\partial{\bf I}}\cdot {\partial H_1\over\partial{\bf w}} =
  0,
\end{equation}
where the subscripts $o$ and $1$ denote the background and first-order
perturbed quantities and $H$ is the Hamiltonian.  The LCBE has been
written in action-angle variables.

I will solve the LCBE by a Laplace transform in time and a Fourier
transform in angle variables (Tremaine \& Weinberg 1984, Weinberg
1989\nocite{TrWe:84,Wein:89}). Let us denote the Laplace transform of
some quantity $q(t)$ by ${\hat q}(s)$ and Fourier transform of $q$ by
$q_{\bf l}({\bf I})$.  The vector quantity ${\bf l}$ is the triple of
`quantum' numbers defining the discrete Fourier series in angle
variables.  Recalling that the frequencies corresponding to the angle
variables are ${\bf\Omega}=\partial H_o/\partial{\bf I}$, the
Fourier-Laplace transform of the LCBE (eq.  \ref{eq:lcbe}) becomes
\begin{equation}
  s{\hat f}_1 + i{\bf l}\cdot{\bf\Omega}{\hat f}_1 - i{\bf l}\cdot{\partial
  f_o\over\partial{\bf I}}{\hat H}_{1\,{\bf l}} = 0.
\end{equation}
Solving for ${\hat f}_1$ yields
\begin{equation}
  \label{eq:flf1}
  {\hat f}_1 = 
    i{\bf l}\cdot{\partial f_o\over\partial{\bf I}}
    {\displaystyle
      {\hat H}_{1\,{\bf l}}
      \over
      s + i{\bf l}\cdot{\bf\Omega}
      }.
\end{equation}

\subsection{Dressed particles}
\label{sec:dress}

We will use the solution in equation (\ref{eq:flf1}) to derive the
response of the continuum stellar system to point particles.  The
effect of a point particle on the system is then the combined effect
of the potential due to point particle and its response.  This has
been called a {\it dressed} point particle by Rostoker \& Rosenbluth
(1960\nocite{RoRo:60}) who first derived the properties of a plasma of
dressed particles.

Following Weinberg (1989\nocite{Wein:89}), I will expand the
perturbation in a biorthogonal series whose basis is constructed from
eigenfunctions of the Laplacian.  The potential or density, then,
trivially follow from the expansion coefficients and the
potential-density pairs, $u_i, d_i$, solve the Poisson equation
explicitly.  I will deviate from traditional notation and define the
true space density corresponding to $u_i$ to be ${\bar d}_i =
-d_i/4\pi G$.  The biorthogonal expansion, then, is in potential and
$-4\pi G$ times the density.  This leads to the convenient
orthogonality condition
\begin{equation}
  \label{eq:ortho} \int_0^L dr r^2 u^{lm\,\ast}_i d^{lm}_j =
  \delta_{ij}.
\end{equation}
The upper limit of the integral in equation (\ref{eq:ortho}) may be
infinite, depending on the pairs.  I will set $G=1$ throughout.

Using this, the spherical harmonic expansion coefficients for a point
mass of mass ${\bar m}=M/N$ at ${\bf r}_o={\bf r}_o(t)$ is
\begin{eqnarray}
  b^{lm}_j &=& -4\pi{\bar m}\int d^3r^\prime
  Y_{lm}^\ast(\theta^\prime,\phi^\prime) u^{lm\,\ast}_j(r^\prime)
  \delta^3({\bf r}^\prime -  {\bf r}_o) \nonumber \\
  &=& -4\pi{\bar m} Y_{lm}^\ast(\theta_o,\phi_o) u^{lm\,\ast}_j(r_o).
  \label{eq:bcoef}
\end{eqnarray}
The gravitational potential of the point mass, the perturbing
Hamiltonian, is then
\begin{equation}
  \label{eq:H_1}
  H_1 = \sum_{l,m}\sum_j b^{lm}_j Y_{lm}(\theta, \phi) u^{lm}_j(r).
\end{equation}

\subsection{Expansion in action-angle series}
\label{sec:actang}

To solve the LBCE using equation (\ref{eq:flf1}), we now need to
expand functions of the form $Y_{lm}(\theta, \phi)g(r)$ in
action-angle variables.  Specifically, for a spherical system, it is
easy to write down the description of a particle orbit in the orbital
plane.  From this, we may derive the general harmonic expansion
following the technique presented in Tremaine \& Weinberg
(1984\nocite{TrWe:84}).  This yields
\begin{eqnarray}
  Y_{lm}(\theta, \phi)g(r) &=& \delta_{m\,l_3} 
  \sum^\infty_{l_1=-\infty}\sum^l_{l_2=-l} i^{m+l_2}
  e^{i{\bf l}\cdot{\bf w}} \times \nonumber \\ &&
  Y_{ll_2}(\pi/2,0) r^l_{l_2m}(\beta)
  W^{l_1}_{l\,l_2\,m}({\bf I})
  \label{eq:ylmexp}
\end{eqnarray}
where $\beta$ is the elevation of the orbital plane defined by the
actions ${\bf I}$, $W$ is
\begin{equation}
  W^{l_1}_{l\,l_2\,m}({\bf I}) = {1\over\pi} \int_0^\pi dw_1\,
  \cos(l_1w_1 +l_2f(w_1))g(r(w_1))
  \label{eq:w1trans}
\end{equation}
with $f(w_1)\equiv\psi-w_2$, and $r^l_{l_2m}(\beta)$ is rotation
matrix for spherical harmonics (e.g. Edmonds 1960\nocite{Edmo:60}).

\subsection{The Laplace transform of $H_{1{\bf l}}$}

Putting the results of \S\S\ref{sec:cbe}--\ref{sec:actang} together,
we can derive the Laplace transform of $H_{1{\bf l}}$ and using
equations (\ref{eq:bcoef})---(\ref{eq:w1trans}) to get the
distribution function of the dressed particle.

First, for a particular spherical harmonic, the action-angle transform
of $H_1$ is
\begin{eqnarray}
  H_{1{\bf l}} &=& {1\over(2\pi)^3}\int d^3w\, e^{-i{\bf l}\cdot{\bf w}}
  \sum_j b^{lm}_j Y_{lm}(\theta,\phi) u^{lm}_j(r) \nonumber \\
  &=& \sum_j Y_{ll_2}(\pi/2,0)r^l_{l_2m}(\beta)W^{l_1\,j}_{ll_2m}({\bf
    I}) b^{lm}_j(t)
\label{eq:H_1l}
\end{eqnarray}
where $b^{lm}_j(t)$ is a some general well-behaved function of time.
The quantity $W^{l_1\,j}_{ll_2 m}({\bf I})$ is shorthand for $W$ with
$g(r) = u^{lm}_j(r)$.  Because the motion of a star on a regular orbit
is quasi-periodic, we will consider terms $b^{lm}_j(t)$ with pure
sinusoidal dependence: $b^{lm}_j(t)=b^{lm}_j\exp(i\omega t)$.
Substituting this into equation (\ref{eq:H_1l}) and Laplace
transforming gives the desired result:
\begin{equation}
  \label{eq:H_1lf}      
  {\hat H}_{1{\bf l}}(s)
  = \sum_j  i^{m+l_2} 
  Y_{ll_2}(\pi/2,0)r^l_{l_2m}(\beta)W^{l_1\,j}_{ll_2m}({\bf I}) 
  {\hat b^{lm}_j(s)},
\label{eq:LH_1l}
\end{equation}
where the Laplace transform of $b^{lm}_j(t)=b^{lm}_j\exp(i\omega t)$
is
\begin{equation}
        {\hat b}^{lm}(s) = {b^{lm}_j\over s-i\omega}.
\end{equation}

\subsection{The response of the dressed particle}

We can compute the density and potential response corresponding to the
dressed particle by integrating the perturbed distribution function
from equation (\ref{eq:flf1}) over velocities:
\begin{equation}
        {\hat\rho}_{\bf l}(s) = \int d^3v {\hat f}_1.
\end{equation}
The Laplace transformed expansion coefficients of the biorthogonal basis
are then
\begin{equation}
        {\hat a}^i = -4\pi\int d^3r Y^\ast_{lm}(\theta,\phi)
        u^{lm\ast}_i(r) \int d^3v {\hat f}_1.
\end{equation}  
This computation is easily performed by noting that the Jacobian of
the canonical transform of $d^3rd^3v$ is unity and we are free to
choose any set.  This procedure is the motivation behind the
biorthogonal expansion.  If we choose $d^3rd^3v = d^3Id^3w$ and use
equation (\ref{eq:ylmexp}), we can do the angle integration trivially.
Then, noting that $d^3I=dE dJ J d(\cos\beta)/\Omega_1(E,J)$, we can do
the integral in $\beta$ using the orthogonality of the rotation
matrices: 
\begin{equation}
\int d\beta \sin(\beta) r^l_{\mu i}(\beta)\, r^l_{\nu
j}(\beta) = {2\over 2l+1}\delta_{\mu\nu}\delta_{ij}
\end{equation}
(Edmonds 1960).  We get:
\begin{eqnarray}
        {\hat a}^{lm}_i &=& -4\pi (2\pi)^3 {2\over 2l+1}
        \sum_j\sum_{\bf l}
        \int {dE\,dJ\,J\over\Omega_1(E,J)} i{\bf l}\cdot{\partial
        f_o\over\partial{\bf I}} 
        \times  \nonumber \\
        && {1\over s+i{\bf l}\cdot{\bf\Omega}} |Y_{ll_2}(\pi/2,0)|^2 
        W_{ll_2m}^{l_1i\ast}({\bf I}) W_{ll_2m}^{l_1j}({\bf I})
        {\hat b}^{lm}_j(s). \nonumber \\
\label{eq:resp1}
\end{eqnarray}

Equation (\ref{eq:resp1}) has the form ${\bf{\hat a}}^{lm} = {\bf{\cal
    R}}^{lm} {\bf{\hat b}}^{lm}$ and describes the response of the
stellar system, ${\hat a}^{lm}$, to the perturbation, ${\hat b}^{lm}$.
The self-gravitating response, then, is the solution to ${\bf{\hat
    a}}^{lm}(s) = {\cal R}^{lm}(s)\cdot\left[ {\bf{\hat a}}^{lm}(s) +
  {\bf{\hat b}}^{lm}(s) \right]$:
\begin{eqnarray}
        {\bf{\hat a}}^{lm}(s) &=& {\bf{\cal D}}^{-1\,lm}(s){\cal
        R}^{lm}(s)\cdot{\bf{\hat b}}^{lm}(s)
         \label{eq:self_resp} \\
        \noalign{\hbox to \hsize{where\hfill}}
        {\bf{\cal D}}^{lm}(s) &\equiv& {\bf{\cal I}} - {\bf{\cal
        R}}^{lm}(s).
\end{eqnarray}
If ${\bf b}=0$, the equation for the response becomes an eigenvalue
problem.  The eigenvalues $s$ are the zeros of the dispersion relation
$\det{\bf{\cal D}}^{lm}(s) = 0$.  For the general stable spherical
stellar system, there will be no modes for $\Re(s)\ge0$ and only in
special cases with restricted phase space does one find oscillatory
modes, $\Re(s)=0$

To evaluate the coefficients as a function of time, we perform the
inverse Laplace transform, deforming the contour to the
$\Re(s)\rightarrow-\infty$:
\begin{eqnarray}
  {\bf a}^{lm}(t) &=& {1\over2\pi i}\int^{c+i\infty}_{c-i\infty}
  ds e^{st} {\bf{\cal D}}^{-1\,lm}(s){\cal R}^{lm}(s)\cdot{\bf{\hat b}}^{lm}
  \nonumber\\
  &=& {1\over2\pi i}\int^{c+i\infty}_{c-i\infty}
  ds e^{st} {\bf{\cal D}}^{-1\,lm}(s){\cal R}^{lm}(s)\cdot{\bf b}^{lm} {1\over
    s-i\omega}.
  \nonumber\\
  \label{eq:deform}
\end{eqnarray}
Assuming that the background is stable with no oscillatory modes,
${\bf{\cal D}}^{-1\,lm}(s)$ has no poles in the half plane
$\Re(s)\ge0$.  The final term gives a pole at $s=i\omega$.  Although
there may be poles in the half plane $\Re(s)<0$, these will vanish for
$t\rightarrow\infty$ relative to the pure imaginary contribution.  To
perform the integral, one may take the $s$ integration into the
phase-space integral for the elements of ${\cal R}^{lm}(s)$.  In
addition to ${\bf{\cal D}}^{-1\,lm}(s)$, the $s$-dependence is in two
simple poles and one finds an integral of the form
\begin{eqnarray}
  {1\over2\pi i}\int^{c+i\infty}_{c-i\infty}
  ds \, e^{st} \, {\bf{\cal D}}^{-1\,lm}(s)
  {1\over s+i{\bf l}\cdot{\bf\Omega({\bf I})}}
  {1\over s-i\omega} = \nonumber \\
  \qquad
  \left[
    {
      {\bf{\cal D}}^{-1\,lm}(i\omega)
      e^{i\omega t} -
      {\bf{\cal D}}^{-1\,lm}(-i{\bf l}\cdot{\bf\Omega}({\bf I}))
      e^{-i{\bf l}\cdot{\bf\Omega}({\bf I})t}
      \over i\omega + i{\bf l}\cdot{\bf\Omega}({\bf I})
      }
  \right]. 
\end{eqnarray}
For large values of $t$, this expression oscillates rapidly and we may
extract the dominant coherent contribution.  There are two cases:
without and with a resonance in the phase space.  The existence of a
resonance in phase space is defined by $\omega + {\bf
  l}\cdot{\bf\Omega}({\bf I})=0$ for $f({\bf I})\not=0$.  For the
non-resonant case, the integrand has no singularity and we can
consider each term separately.  The first term yields a contribution
in phase with the perturbation while the second term in the brackets
oscillates incoherently and makes no net contribution.  The second
term, therefore, can be ignored.  For the resonant case, the
contribution at large $t$ has a sharp peak about $\omega + {\bf
  l}\cdot{\bf\Omega}({\bf I})=0$ as $t\rightarrow\infty$.  Expanding
${\bf{\cal D}}^{-1\,lm}(i\omega)$ about $i\omega$ and retaining only
dominant terms, one finds the contribution near the resonance is
\begin{eqnarray}
  \left[\quad\right] &\approx& {\bf{\cal D}}^{-1\,lm}(i\omega)
  e^{i(\omega - {\bf l}\cdot{\bf\Omega}({\bf I}))t/2}\ 2\,
  {
    \sin\left[{\omega 
        + {\bf l}\cdot{\bf\Omega}({\bf I})\over2}t\right]
    \over 
    \omega + {\bf l}\cdot{\bf\Omega}({\bf I})
    } \nonumber \\
  \noalign{\leftline{or}}
  \lim_{t\rightarrow\infty} \left[\quad\right] &=& 
  {\bf{\cal D}}^{-1\,lm}(i\omega)
  e^{i\omega t} 2\pi \delta\left(
    \omega + {\bf l}\cdot{\bf\Omega}({\bf I})
  \right).
\end{eqnarray}
We will adopt the latter asymptotic form here and see in the final
computation that the $\exp(i\omega t)$ will cancel leaving only the
delta functions.  Putting both cases together yields a simple
expression
\begin{eqnarray}
        \label{eq:dcoef}
        a^{lm}_i(t) &=& e^{i\omega t} \sum_\nu {\cal
          M}^{lm}_{i\nu}(\omega)\, b^{lm}_\nu, \\
        {\cal M}^{lm}_{i\nu} &\equiv& -4\pi (2\pi)^3 {2\over 2l+1}
        \sum_j{\cal D}^{-1\,lm}_{ij} (i\omega) \sum_{{\bf l}}
        \nonumber \\ &&
        \int {dE\,dJ\,J\over\Omega_1(E,J)} {\bf l}\cdot{\partial
        f_o\over\partial{\bf I}} \times
        \nonumber \\
        && |Y_{ll_2}(\pi/2,0)|^2\,
        W_{ll_2m}^{l_1j\ast}({\bf I}) W_{ll_2m}^{l_1\nu}({\bf I})
        \times \nonumber \\
        && \left[
          {\cal P}{1\over\left(\omega + {\bf l}\cdot{\bf\Omega}({\bf
          I})\right)} +
          2\pi i \delta\left(
            \omega + {\bf l}\cdot{\bf\Omega}({\bf I})\right)
          \right]
        \nonumber \\ \label{eq:mik}
\end{eqnarray}
To simplify notation, we have explicitly noted in equations
(\ref{eq:dcoef}) and (\ref{eq:mik}) that the solution takes the the
form ${\bf a}^{lm}(t) = \exp(i\omega t){\bf{\cal
    M}}^{lm}(\omega)\cdot{\bf b}^{lm}$.  The integrals in the matrix
elements of ${\cal D}^{lm}$ may be analytically continued using the
Landau prescription (e.g. Krall \& Trivelpiece 1973\nocite{KrTr:73})
after a conformal mapping of the discrete interval in $E$.  Notice
that this expression takes both resonant and non-resonant cases into
account; without a resonance, the delta function does not contribute
and principal value of a non-singular integrand is the integrand
itself.  One recovers the non-self-gravitating but global response by
setting ${\cal D}^{lm}_{ij} = \delta_{ij}$.

\subsection{Energy in fluctuations}

We will now use equation (\ref{eq:dcoef}) to evaluate the fluctuation
energy at different spatial scales assuming that individual particles
are uncorrelated.  The particle wakes do in fact give rise to
correlations but this is of higher order in $1/N$ in the BBGKY
expansion (cf. Gilbert 1969\nocite{Gibe:68}) than the lowest-order
effect we will consider here.

This leaves us with individual particles reacting coherently to the
effect of their own wakes.  Because the particles are uncorrelated,
the number density of particles at ${\bf I}_o, {\bf w}_o$ at time $0$
and at ${\bf I}^\prime, {\bf w}^\prime$ at time $t$ is
\begin{eqnarray}
        {\cal P}({\bf I}_o, {\bf w}_o, 0; {\bf I}^\prime, {\bf
        w}^\prime, t) &=& {\cal P}({\bf I}_o, {\bf w}_o)
        \delta({\bf I}^\prime - {\bf I}_o)
        \times \nonumber \\ &&
        \delta({\bf w}_o - {\bf w}^\prime + {\bf\Omega}({\bf I}_o)t)
        \nonumber \\
        \label{eq:dist2}
\end{eqnarray}
where ${\cal P}({\bf I}_o, {\bf w}_o)$ is the equilibrium particle
distribution with
\begin{equation}
N=\int d^3I_o d^3w_o {\cal P}({\bf I}_o, {\bf w}_o).
\end{equation}
Direct substitution demonstrates that equation (\ref{eq:dist2}) solves
the Liouville equation with the initial condition ${\bf I}^\prime =
{\bf I}_o$ and ${\bf w}^\prime = {\bf w}_o$ at $t=0$.  Similarly,
integrating equation (\ref{eq:dist2}) over all coordinates gives $N$.

For a given harmonic $lm$, the fluctuation energy is then
\begin{eqnarray}
        \label{eq:fenergy}
        {1\over2}\langle\Phi_1\rho_1\rangle &=& {1\over2}
        \int d^3r \langle
        \sum_i a^{lm\ast}_i Y_{lm}^\ast(\theta,\phi) u^{lm\,\ast}_i(r)
        \times \nonumber \\ &&
        {1\over4\pi}\sum_k a^{lm}_k Y_{lm}(\theta,\phi) d^{lm}_k(r) \rangle
        \nonumber \\
        &=&
        {1\over 2}\left(-{1\over4\pi}\right)
        \langle a^{lm\ast}_i a^{lm}_i\rangle
\end{eqnarray}
where the expectation value of some quantity $\Xi$ is defined by
\begin{eqnarray}
        \label{eq:l2def}
        \langle\Xi\rangle &\equiv& 
                \int d^3I_o d^3w_o d^3I^\prime d^3w^\prime
                {\cal P}({\bf I}_o, {\bf w}_o, 0; {\bf I}^\prime, {\bf
                w}^\prime, t) \,\Xi \nonumber \\
        &=&
                N \int d^3I_o d^3w_o d^3I^\prime d^3w^\prime
                f_o({\bf I}_o)
        \delta({\bf I}^\prime - {\bf I}_o)
        \times \nonumber \\ &&
        \delta({\bf w}_o - {\bf w}^\prime + {\bf\Omega}({\bf I}_o)t) \,\Xi.
        \nonumber \\
\end{eqnarray}

Applying equations (\ref{eq:dcoef}) and
(\ref{eq:l2def}) to equation (\ref{eq:fenergy}) gives
\begin{eqnarray}
        \label{eq:fenergy2}
        {1\over2}\langle\Phi_1\rho_1\rangle &=& -{4\pi{\bar m}\over2}
        (2\pi)^3 \left({2\over2l+1}\right)
        \sum_{{\bf l}} \sum_{i\mu\nu} \times
        \nonumber \\
        &&
        \int {dE\,dJ\,J\over\Omega_1(E,J)}
        |Y_{ll_2}(\pi/2,0)|^2 f_o(E, J) \times
        \nonumber \\
        &&
        {\cal M}^{lm\,\ast}_{i\mu}({\bf l}\cdot{\bf\Omega}({\bf I}))
        {\cal M}^{lm}_{i\nu}({\bf l}\cdot{\bf\Omega}({\bf I})) \times
        \nonumber \\
        &&
        W_{ll_2m}^{l_1\mu\ast}({\bf I}) W_{ll_2m}^{l_1\nu}({\bf I}).
        \nonumber \\
\end{eqnarray}
Gathering terms, this can be simplified as follows:
\begin{eqnarray}
  \label{eq:fenergy3}
  {1\over2}\langle\Phi_1\rho_1\rangle
  &=& -{4\pi{\bar m}\over2}
  (2\pi)^3 \left({2\over2l+1}\right) \sum_{{\bf l}} \sum_i \times
  \nonumber \\
  &&
  \int {dE\,dJ\,J\over\Omega_1(E,J)}
  |Y_{ll_2}(\pi/2,0)|^2 f_o(E, J)
  \nonumber \\
  &&
  \left|\sum_\nu
    {\cal M}^{lm}_{i\nu}({\bf l}\cdot{\bf\Omega}({\bf I}))
    W_{ll_2m}^{l_1\nu}({\bf I})\right|^2.
  \nonumber \\
\end{eqnarray}
Note that each term in the fluctuation energy, equation
(\ref{eq:fenergy3}), is negative definite as expected.  The
contribution for each triple in the angle expansion, ${\bf l}$, and
each term in the basis expansion $k$ may be tabulated separately.

We may compute the fluctuation energy in the absence of gravity by
returning to equation (\ref{eq:fenergy}) and evaluating $\langle
a^{lm\ast}_i a^{lm}_i\rangle$ without any dynamics.  For $N$
particles, the sample value for $a_i^{lm}$ is $a_i^{lm} = -{1/N}\sum_k
4\pi{\bar m} Y_{lm}^\ast(\theta_k,\phi_k) u^{lm\,\ast}_j(r_k).$ Using
the expectation defined by equation (\ref{eq:l2def}) one finds:
\begin{equation}
  \label{eq:fenergy4}
  {1\over2}\langle\Phi_1\rho_1\rangle
  = -{4\pi{\bar m}\over2} \sum_i \int d^3r\, \rho({\bf r})
  u^{lm\,\ast}_i(r)
  u^{lm}_i(r).
\end{equation}
This is identical to equation (\ref{eq:fenergy3}) with out the
particle dressing: ${\cal M}^{lm}_{i\nu} = \delta_{i\nu}$.

\section{Examples of fluctuation spectrum standard models}
\label{sec:examples}

\begin{figure}
\mbox{\epsfxsize=\colsize\epsfbox{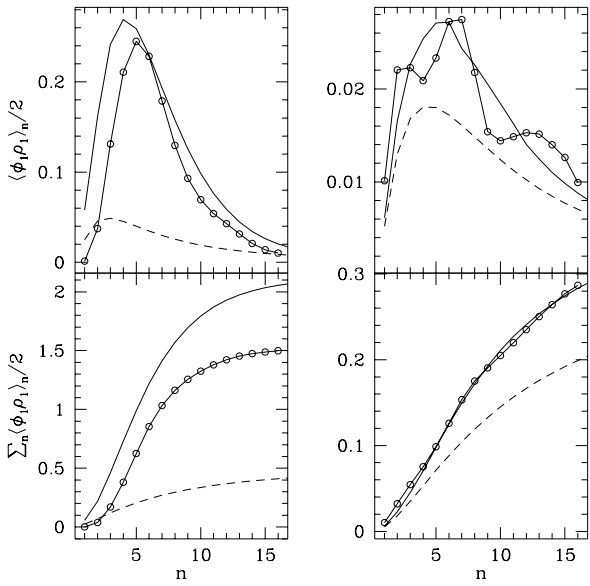}}
\caption{Top row: Fluctuation energy with orthogonal function index in
  units of background potential energy.  Bottom row: Cumulative
  fluctuation energy with orthogonal function indices less than $n$.
  The left and right columns show the $l=m=1$ and $l=m=2$ harmonics
  respectively. The spatial profile of the potential functions is
  shown in Fig. \protect{\ref{fig:bessel}} for $l=1, 2$.  The solid
  and dashed line follows from the analytic calculation for the
  self-gravitating response and Poisson fluctuations alone,
  respectively.  The open circles are computed from the time-averaged
  expansion coefficients in the n-body simulation.}
\label{fig:energy}
\end{figure}

\begin{figure}
\mbox{\epsfxsize=\colsize\epsfbox{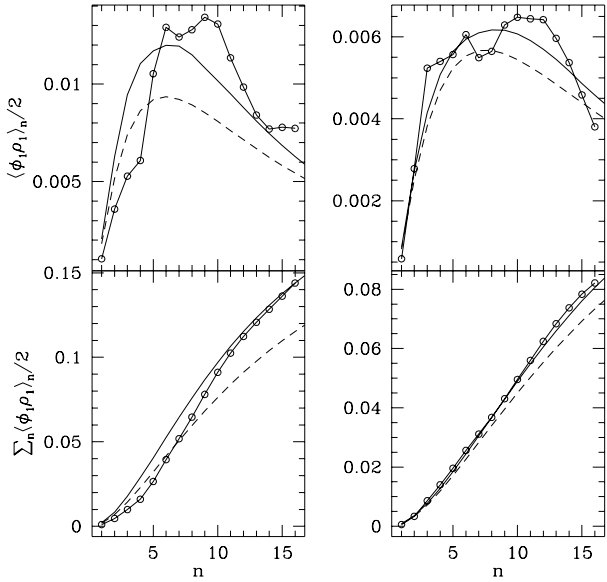}}
\caption{As in Fig. \protect{\ref{fig:energy}} but 
  for the $l=m=3$ (left) and $l=m=4$ (right) harmonics.  }
\label{fig:energy2}
\end{figure}

\begin{figure}
\mbox{\epsfxsize=\colsize\epsfbox{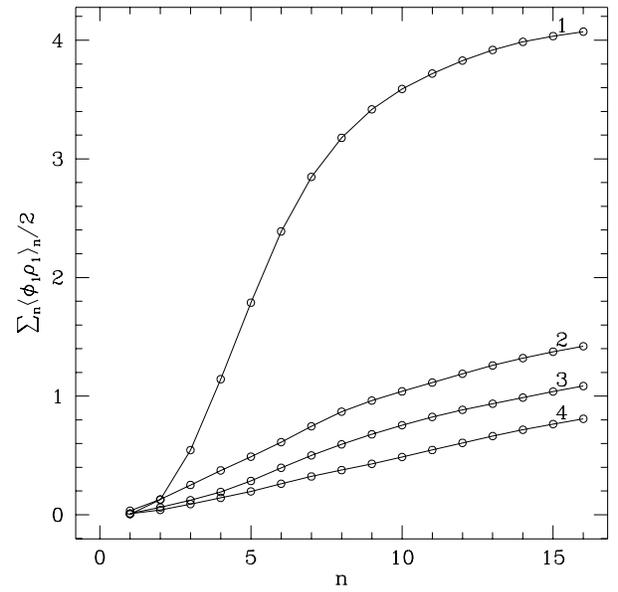}}
\caption{Cumulative fluctuation energy for all harmonics of given $l$
  (labeled)
  with orthogonal function indices less than $n$.  Shown in units of
  background potential energy as in Fig. \protect{\ref{fig:energy}}.
  }
\label{fig:engytot}
\end{figure}

\begin{figure*}
\mbox{
\mbox{\epsfxsize=\colsize\epsfbox{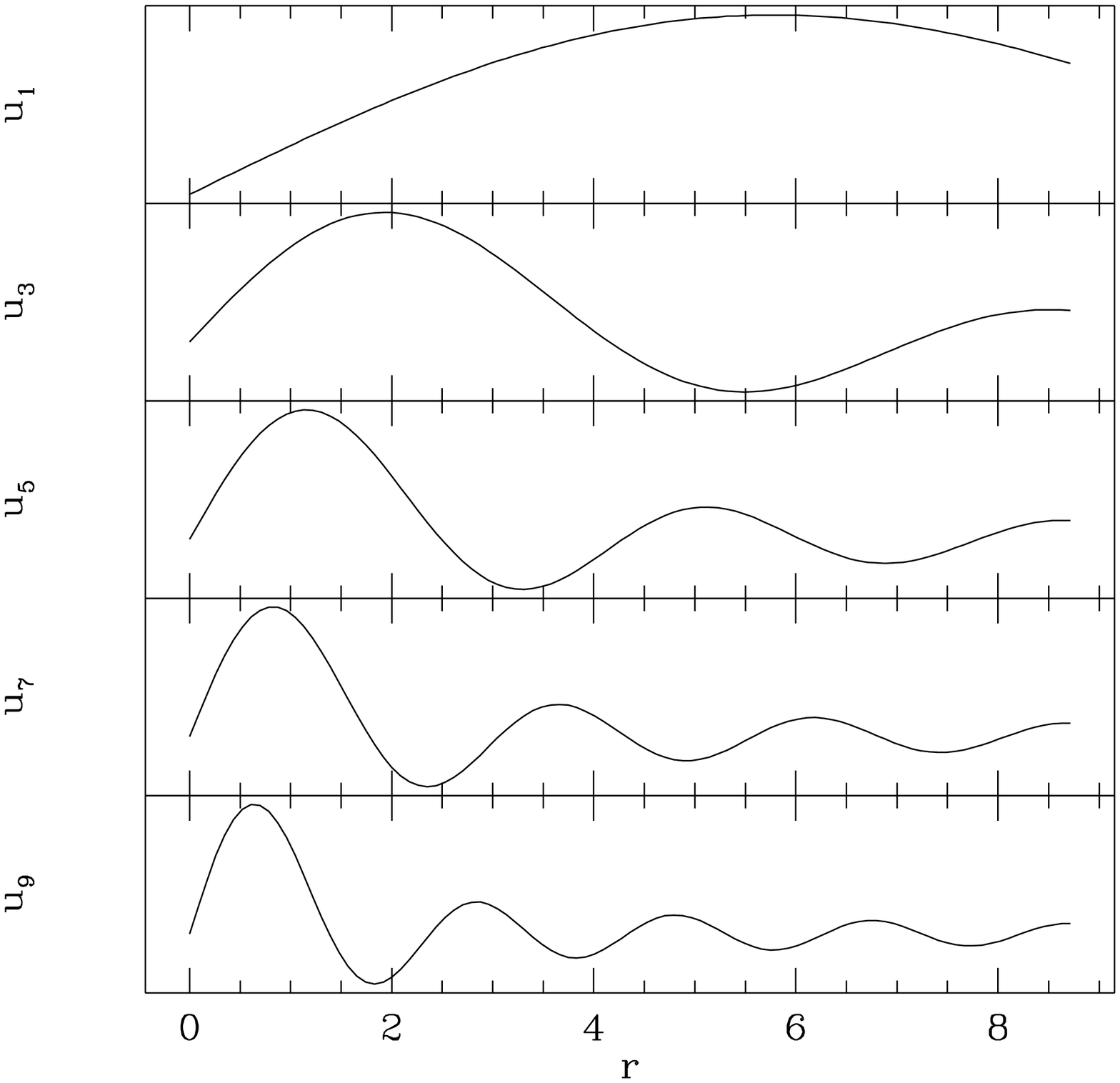}}
\mbox{\epsfxsize=\colsize\epsfbox{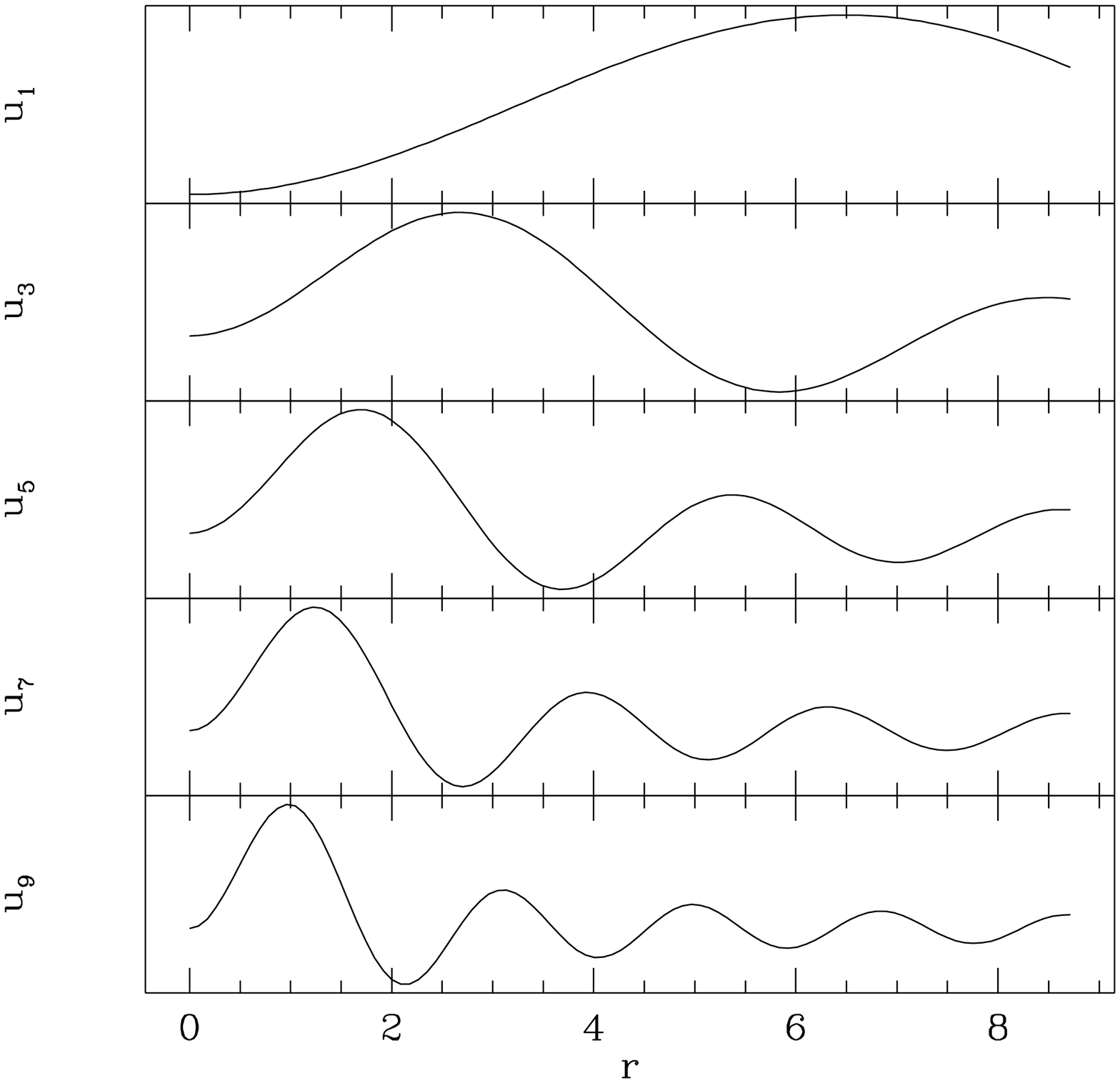}}
}
\caption{Potential functions used in fluctuation energy calculation in
Fig. \protect{\ref{fig:energy}}.}
\label{fig:bessel}
\end{figure*}

We will look at both King models and core-free Hernquist models.
First, we apply equation (\ref{eq:fenergy3}) to a King model ($W_0=5$)
for harmonics $l=m=1, 2, 3, 4$ and compare with a SCF simulation for
100,000 particles and $l_{max}=4$.  Figures \ref{fig:energy} and
\ref{fig:energy2} show the fluctuation energy per expansion function
plotted against the index of expansion functions (cf.  Fig.
\ref{fig:bessel}) and the cumulative fluctuation energy less than
given index.  One energy unit is equal to the total gravitational
potential energy of the unperturbed sphere and ${\bar m}=1$.  There
are slight systematic differences between the n-body results and the
predictions but all-in-all, the n-body simulation follows the analytic
predictions fairly well and confirms the excess power at low order due
to amplification by self gravity.  This excess is illustrated in
Figure \ref{fig:engytot} which shows the total energy for harmonic
orders $l=1$--4 derived from the simulation.  For $l=1$, the
enhancement is roughly a factor of 6.  The large magnitude for $l=1$
is due to the stochastic excitation of a weakly-damped mode.  For
$l=2$, the enhancement is roughly a factor of 1.5.  For $l>4$ the
power enhancement due to self gravity is negligible.  The size of
coherent structures decrease with increasing harmonic order and
therefore higher harmonics have power at smaller and smaller scales
for which self gravity is less important.  The index at peak power
increases with $l$ as expected.

\begin{figure}
\mbox{\epsfxsize=\colsize\epsfbox{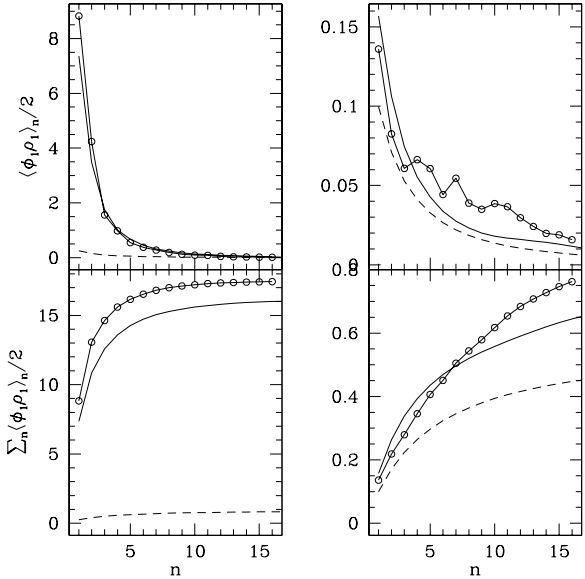}}
\caption{As in Figure \protect{\ref{fig:energy}} but for the Hernquist
  model.}
\label{fig:energyh}
\end{figure}

The case for the core-free Hernquist model is shown in Figure
\ref{fig:energyh} for harmonics $l=m=1, 2$.  The profiles are more
sharply peaked about $n=1$ because the expansion functions well
matched to the model profile (Hernquist \& Ostriker
1992\nocite{HeOs:92}).  Especially for the $l=m=1$ harmonic, the
agreement between the expansion and the simulation is better in this
case.  This is probably due to the choice of expansion functions.  For
the $l=m=2$, the power appears systematically high at larger radial
order.  As for the King model, the power at $l=1$ has the largest
enhancement by self-gravity, now by roughly a factor of 15 and this
amplitude is verified by the simulation.  The enhancement at $l=2$ is
similar, roughly a factor of 1.5.

The root energy indicates the expected magnitude of the density or
potential fluctuation and can be multiplied by $\sqrt{1/N}$ to
estimate the magnitude for an $N$ particle simulation.  For galaxian
disk embedded in a massive halo, an large-scale $0.1\%$ distortion in
the halo can have interesting consequences for disk evolution (cf.
Weinberg 1997).  In order to realistically test dynamical hypotheses
for disk-halo interactions, we need to suppress noise below this
level.  This requires live halos with $N>10^7$ particles for S/N$>10$.

In particular, the fluctuating dipole ($l=1$) force field
differentially accelerates and bends the disk, causing thickening.
The quadrupole ($l=2$) in a arbitrary orientation warps the disk,
causing the sort of warp discussed by Weinberg (1995,
1997\nocite{Wein:95b,Wein:97}) but now due to noise rather than a
satellite wake.  Even without self gravity, discreteness noise is
sufficient to require $N>10^6$.

\section{Conclusions and discussion}
\label{sec:conclu}

This paper considers discreteness noise and its amplification in
n-body simulations with particular attention a halo's effect on an
embedded disk.  The overall conclusions are as follows:
\begin{enumerate}
  \item The SCF n-body algorithm does not suppress Poisson
    fluctuations due to finite number of particles and Poisson
    fluctuations alone can result in significant distortions.
  \item The low-order response is amplified by the coherent
    self-gravitating response of the entire system.
  \item As discussed in Weinberg (1994\nocite{Wein:94}), spherical
    systems have weakly damped modes at harmonics $l=m=1,2$.  These
    are excited by noise and enhance the fluctuation power. This is
    clearly seen in Figures \ref{fig:energy} and \ref{fig:energyh}.
    Detailed agreement between the linearized solution and simulation
    reaffirms the importance of these modes to the self-gravitating
    response.
  \item For $N$ particles, the $l=m=1(2)$ harmonics have roughly $4/N
    (1.5/N)$ times the energy of the background for the $W_0=5$ King
    model and $15/N (1.5/N)$ for the Hernquist model.  The
    amplification does not rely on the existence of a core.
  \item The noise amplitude in an n-body simulation with $N=10^5$ is
    comparable to the the amplitude required to produce a warp
    (Weinberg 1997).  This analysis suggests simulations with
    $N\gtrsim10^7$ will be necessary to achieve healthy a signal to
    noise ratio.  Conversely, a halo with $N=10^5$ corresponds to
    masses of roughly 2 to $6\times10^6\msun$.  This is comparable to
    the black hole mass required to produce the disk scale height by
    the Lacey \& Ostriker (1985\nocite{LaOs:85}) mechanism.
    Fluctuations at this level are likely to excite bending modes in
    addition to increasing the disk scale height.
  \item Although the fluctuations described in Figure \ref{fig:energy}
    include self-gravity, self-gravity is a small ($<10\%$)
    perturbation for harmonics $l>4$.
\end{enumerate}

The gravitationally amplified noise described here is unlikely to be a
significant contribution to the overall evolution of a relaxing system,
such as a globular cluster.  However, the noise-excited $l=1$
structure may be sufficient to offset the cluster halo from its core.
This could decrease the lifetime of any loss-cone population and
subsequently decrease the rate of core cooling.

This calculation investigates the magnitude and nature of halo
fluctuations assuming that the halo is an equilibrium phased-mixed
distribution.  However, the outer parts of galaxies can barely count
10 dynamical times so inhomogeneities will not have had time to phase
mix and continued disturbance from mergers will not have relaxed (see
Tremaine 1992\nocite{Trem:92} for additional discussion).  It is
likely that intrinsic fluctuations may play a significant role in
long-term galaxian evolution.  In particular, a companion paper
(Weinberg 1997) elaborates the nature of satellite halo excitation and
its role in producing warps using a similar formalism.  N-body
simulation of this process revealed that intrinsic noise had a similar
effect on the disk.  It is difficult to deduce the power produced by
phase mixing streams based on the point-mass noise from $10^5$, $10^6$
or $10^7M_\odot$ fuzzy blobs and the degree of granularity will depend
on the nature of the dark matter and halo formation history.
Speculating further nonetheless, it is conceivable that a wide variety
of observed disk structure such as arms and bars may have its roots in
halo structure.

\section*{Acknowledgements}
I thank Enrico Vesperini for helpful comments.  This work was
supported in part by NSF grant \# AST-9529328 and the Sloan
Foundation.

\bsp
\label{lastpage}
\end{document}